\documentstyle[12pt]{article}
\begin{document}
Nowadays it is well-known that the Lorenz model is a paradigm one for low 
dimensional chaos in dynamical systems in synergetics and this model or
its modifications are widely investigated in connection with modelling 
purposes in
meteorology, hydrodynamics,
laser physics,superconductivity,electronics,oil industry
etc.,see,e.g.refs.1-15 and references therein. 
From the mathematical
point of view, the Lorenz model is a system of nonlinear equations.Needless
to say that in general it is vertually impossible to find a closed 
analytical solutions to the  most of nonlinear equations.So one should take 
advantage of asymptotic approaches or have to recourse to the help of 
numerical simulations, which  is not comprehensive for multi-parameter 
systems.In this paper we apply the asymptotic method for
singularly perturbed nonlinear systems ( ref.16 and references therein) 
to the Lorenz model.Earlier this method was applied by the author of this
paper in refs.17-22 and references therein.\\ 
The system under study is of the form:
$$\frac{dx}{dt}=\sigma (y-x),$$
$$\frac{dy}{dt}=rx-y-xz,\hspace*{0.7cm} (10)$$
$$\frac{dz}{dt}=xy-bz,$$
where $x $,$y$ and $z$ are dynamical variables;$\sigma$ ,$r$ and $b$ are the 
parameters of the sytem (1).In general initial conditions are :
$x(t=0)=x(0)$, $y(t=0)=y(0)$, $z(t=0)=z(0)$. Here we deliberately 
don't define the physical meaning of the dynamical variables and parameters,
as in different fields of science the meaning is different.
The system (1) will be investigated in the extreme cases, when $\sigma^{-1}$
and $b^{-1}$ are the small parameters of the problem.According to literature
such values of parameters are quite possible (values $b>>1$ 
are possible for non-laser systems ref.23).In order to apply the 
above-mentioned asymptotic method we rewrite the first equation of system (1) 
in the following form
$$\sigma^{-1}\frac{dx}{dt} =y-x, \hspace*{3cm}(2)$$
According to the theory in the zero-th order of $\sigma^{-1}$ the 
solution of the system in the larger time domain is determined by the 
so-called reduced system :
$$y (t)=x(t),$$
$$\frac{dy}{dt}=rx-y-xz,\hspace*{0.7cm} (3)$$
$$\frac{dz}{dt}=xy-bz,$$
with the initial conditions $y(0),z(0)$.In order to find the solution of the
system (1) for small time domain (in the so-called boundary layer) we should
make transition to the ``new'' time $\tau = \sigma t$,in other words the time 
domain $t=0$ is to be seen through microscope (as it is magnified $\sigma$ 
times).After this operation for the solution of the system (1)in the boundary 
layer in the zero-th order of $\sigma$ we obtain easily:
$$y(t)=y(0), z(t)=z(0),$$
$$x(t)=x(0)\exp (-\sigma t) + y(0)(1-\exp (-\sigma t)),\hspace*{2cm}(4)$$
The solution of the system (1) in the whole time domain is to be 
constructed by linking (4) and solution to the system (3).The system (2)
is the another nonlinear system to be studied by the asymptotic method in 
question.For these purposes we should make the following 
tranformations in the system (3):$y=b^{\frac{1}{2}}y_{1}$, $z=z_{1}$ and 
after that rewrite the third equation of the nonlinear system (3) in the 
following form:
$$b^{-1}\frac{dx_{1}}{dt} =y_{1}^{2}-z_{1},\hspace*{2cm}(5)$$
In other words system (3) will be studied on condition that $b^{-1}$is the 
small parameter.Acting as in the case of initial system (1) in the zero-th 
order of $b^{-1}$ we easily obtain the solution of the system (3) in the
whole time domain:
$$y(t)=b^{\frac{1}{2}}y_{1},$$
$$z(t)=y_{1}^{2} + (z_{1}(0)-y_{1}^{2}(0))\exp (-bt), \hspace*{1cm}(6)$$
If $r$ is not equal to unity, then
$$y_{1}^{2}(t)=(r-1)\exp (2(r-1)t)(A+ \exp (2(r-1)t))^{-1},\hspace*{1cm}(7)$$
If $r=1$ then 
$$y_{1}^{2}(t)= (y_{1}^{2}(0) + 2t)^{-1},\hspace*{2cm}(8)$$
In formulai (7) and (8)
$$A=((r-1)-y_{1}^{2}(0))y_{1}^{2}(0), y_{1}(0)=b^{-\frac{1}{2}}y(0),\hspace*{2cm}(9)$$
Thus for the solution of the initial system (1) in the zero-th order
of $\sigma ^{-1}$, $b^{-1}$ in the whole time domain we obtain:
$$x(t)=(x(0)-y(0))\exp (-\sigma t) + y_{1}(t)b^{\frac{1}{2}},$$
$$z(t)=y_{1}^{2}(t) + (z_{1}(0)- y_{1}^{2}(0))\exp (-bt),$$
$$y(t)=y_{1}(t)b^{\frac{1}{2}},\hspace*{2cm}(10)$$
As the analysis of equations (10) show the characteristic time $t^{charact}$
of changing $y(t)$ from $y(0)$ to the stationary value of $y$, $y^{stat}$
is of the order of $\vert(r-1)\vert$, if $r$ is not equal to unity; if
$r=1$, then $t^{charact}\approx y_{1}^{-2}(0)$.Changing of $z(t)$ from $z(0)$
to $z^{stat}$ occurs through the intermediary quasistationary state 
$z^{qstat}=y_{1}^{2}(0)$ with the required time to achieve this state 
$t^{qstat}=b^{-1}$; transition from $z^{qstat}$ to $z^{stat}$ takes the amount 
of time equal to $t^{charac}$.The changing of $x(t)$ from $x(0)$ to
$x^{stat}$ occurs with the same scenario as for $z(t)$; the only difference
is that the intermediary state for $x(t)$ is $y(0)$ with the transition time
(from $x(0)$) $t^{tr}=\sigma ^{-1}$.\\
In this paper we restricted ourselves to the case of zero-th order 
approximation.In the higher order approach  we encounted with analytically
hard treatable equations.The degree of adequacy of our formulai
could be checked by the comparison with the behavior of the initial Lorenz
model when independent variable $t$ goes to infinity.Before comparing one
should make clear that the asymptotic theory is not applicable when the 
nonlinear system develops full instability ref.4.This is the case for the 
Lorenz model, when for the given values of $\sigma$ and $b$ the value of $r$
exceeds the so-called critical value (onset of chaotic behavior):  
$$r_{cr}=(3+b+\sigma )\sigma (\sigma -1-b)^{-1},\hspace*{2cm}(11)$$
At $r>r_{cr}$ the non-zero fixed points (or steady states) of the Lorenz 
system 
$$ x^{stat}=y^{stat}=\pm (b(r-1))^{\frac{1}{2}},$$
$$z^{stat}=r-1,\hspace*{2cm}(11)$$ 
become unstable,and there is a strange attractor over which a chaotic motion 
takes place. For $\sigma =10$, $b=\frac{8}{3}$ the critical number is equal 
to $r_{cr}=24.74$.Also it is known that the partial loss of instability
in the Lorenz model occurs when $r>1$:with this value of $r$ the trivial 
steady state loses its stability ref.4.\\
The analyses of our formulai show that indeed when $r>1$ the system goes 
to the nontrivial steady state.In the contrary case the trivial steady 
state is obtained.Of course the presented here in this work results 
are of the simplest one for the Lorenz model which is capable to exhibit
highly complicated behavior.But they are adequate at least for some extreme 
cases.\\ 
In conclusion in this work we investigate the Lorenz model in synergetics 
with the asymptotic method for singularly perturbed nonlinear systems for
some limiting cases. The times of achieving quasistationary and 
stationary states are estimated.\\
The author thanks  the JSPS for the Fellowship.

\end{document}